\documentclass[reprint,superscriptaddress, %groupedaddress,%nofootinbib,longbibliography,%nobibnotes,bibnotes,amsmath,amssymb,aps,prb
]{revtex4-2}
\usepackage[utf8]{inputenc}
\usepackage[english]{babel}

\usepackage[plainpages = false, pdfpagelabels, bookmarks, bookmarksopen = true, bookmarksnumbered = true, breaklinks = true, linktocpage, colorlinks = true, linkcolor = blue, urlcolor  = blue, citecolor = blue,  anchorcolor = green, hyperindex = true,  hyperfigures]{hyperref}

\usepackage{enumerate}
\usepackage{tabularx}
\usepackage{makecell, multirow}
\usepackage{tikz}
\usetikzlibrary{patterns}
\usepackage{slashed}
\usepackage{physics}
\usepackage{verbatim}
\usepackage{svg}
\usepackage{mathtools}
\usepackage{cancel}
\usepackage{bbold}
\usepackage{bm}
\usepackage{booktabs}
\usepackage[caption=false,singlelinecheck=off]{subfig}
\usepackage{blindtext}
\usepackage{bbm}
\usepackage{graphicx}
\usepackage{dcolumn}
\usepackage{bm}
\usepackage[normalem]{ulem}
\usepackage{xcolor}

\newcommand{\Shom}{S_\mathrm{HOM}}

\newcommand{\fu}{\phi_\mathrm{u}}
\newcommand{\fl}{\phi_\mathrm{l}}
\newcommand{\fj}{\phi_j}
\newcommand{\psu}{\psi_\mathrm{qp,u}}
\newcommand{\psl}{\psi_\mathrm{qp,l}}
\newcommand{\psj}{\psi_{\mathrm{qp},j}}

\newcommand{\pselj}{\psi_{\mathrm{el},j}}

\newcommand{\rj}{\rho_j}
\newcommand{\tu}{t_\mathrm{u}}
\newcommand{\tl}{t_\mathrm{l}}
\newcommand{\tj}{t_j}
\newcommand{\xu}{x_\mathrm{u}}
\newcommand{\xl}{x_\mathrm{l}}

\newcommand{\Itun}{I_\mathrm{tun}}

\begin{document}
%%%%%%%%%%%%%%%%%%%%%%%%%%%%%%%%%%%%%%%%%%%%%%%%%%%%%%%%%%%%%%%%%%%%%%%%%%%%%%

%TC:ignore

\title{\texorpdfstring{Exchange-phase erasure in anyonic Hong-Ou-Mandel interferometry}{}}
\author{Sushanth Varada}
\affiliation{Department of Microtechnology and Nanoscience (MC2),Chalmers University of Technology, S-412 96 G\"oteborg, Sweden}
\affiliation{Department of Physics and Astronomy, Uppsala University, Box 516, S-751 20 Uppsala, Sweden}
\author{Christian Sp\r{a}nsl\"{a}tt}
 \affiliation{Department of Engineering and Physics, Karlstad University, Karlstad, Sweden}
 \affiliation{Department of Microtechnology and Nanoscience (MC2),Chalmers University of Technology, S-412 96 G\"oteborg, Sweden}
\author{Matteo Acciai}
\affiliation{Scuola Internazionale Superiore di Studi Avanzati (SISSA), Via Bonomea 365, 34136 Trieste, Italy}
\affiliation{Department of Microtechnology and Nanoscience (MC2),Chalmers University of Technology, S-412 96 G\"oteborg, Sweden}

\date{\today}
\begin{abstract}
Two-particle interferometry is an important tool for extracting the exchange statistics of quantum particles. We theoretically investigate the prospects of such interferometry to probe the statistics of point-like anyonic excitations injected in a Hong-Ou-Mandel (HOM) setup based on a quantum point contact device in the fractional quantum Hall regime. We compute the standard HOM ratio, i.e., the ratio of tunneling noises for two- and one-particle injections, and find that for point-like anyons,
it
only depends on the temperature and the anyon scaling dimension. Importantly, the latter is not necessarily related to the exchange phase. In fact, we establish that the HOM ratio does not reveal the exchange phase of the injected anyons: For injection-time delays that are small compared to the thermal time scale, we find that the exchange phase accumulated due to time-domain braiding between injected and thermally activated anyons is erased due to two mutually canceling sub-processes. In contrast, for time delays large compared to the thermal time, only a single sub-process contributes to the braiding, but the accumulated phase is canceled in the HOM ratio. These findings suggest caution when interpreting HOM interferometry experiments with anyons and approaches beyond the standard HOM ratio are thus necessary to extract anyonic statistics with two-particle interferometry experiments.
\end{abstract}
\maketitle
%%%%%%%%%%%%%%%%%%%%%%%%%%%%%%%%%%%%%%%%%%%%%%%%%%%%%%%%%%%%%%%%%%%%%%%%%%%%%%
\textcolor{blue}
{\noindent{\textit{Introduction.---}}}
Quantum exchange statistics is a tenet of modern physics, underpinning phenomena from Bose Einstein condensation to the periodic table of elements to the formation of stars.  In ordinary, three-dimensional space, quantum mechanics predicts~\cite{Laidlaw1971Mar} that elementary particles belong to one of two fundamental types: bosons and fermions. These types correspond to the many-body wavefunction of indistinguishable particles acquiring upon particle exchange a phase factor $e^{i\vartheta}$, with $\vartheta=0$ and $\vartheta=\pi$ for bosons and fermions, respectively. Two-dimensional systems, however, permit particles beyond this dichotomy~\cite{Leinaas1977Jan,Goldin1980Apr}. There, particle exchange can generate \textit{any} phase angle $\vartheta$ and the particles are then referred to as (Abelian) anyons~\cite{Wilczek1982Oct}.

In this work, we investigate the prospects to detect anyons with two-particle interferometry in the fractional quantum Hall (FQH) effect~\cite{Stormer1982,Laughlin1983}. While the charges of FQH quasiparticles were established decades ago~\cite{De-Picciotto1997Sep,Saminadayar1997Sep} to be fractions of the electron charge, it was only in 2020 that the FQH quasiparticles were experimentally established to be anyons~\cite{Bartolomei2020Apr,Nakamura2020Sep}: In Ref.~\cite{Nakamura2020Sep} an anyonic phase angle $\vartheta=\pi/3$ was observed in the FQH state at filling $\nu=1/3$ in a Fabry-P\'erot interferometer in the form of abrupt and reproducible phase jumps in the measured Aharonov-Bohm conductance patterns. This approach was later extended to other fillings in Refs.~\cite{Nakamura2022Jan,Nakamura2023Oct}, and has recently been implemented also in graphene-based devices~\cite{Samuelson2024Mar,Werkmeister2025Apr}. A complementary approach was taken in Ref.~\cite{Bartolomei2020Apr} which reported the impact of $\vartheta$ in the noise signal of a two-particle interferometer, in the so-called collider geometry. In that work, based on an earlier theoretical proposal~\cite{Rosenow2016Apr}, $\vartheta$ was proposed to affect the current correlations generated when two dilute beams of anyons carried by chiral edge states impinge on a beam splitter realized with a quantum point contact (QPC). This observation was later confirmed in additional, independent experiments~\cite{Ruelle2023Mar,Glidic2023Mar,Lee2023May}, and has subsequently spurred several theoretical works~\cite{Morel2022Feb,Lee2020Nov,Lee2022Nov,Idrisov2022Aug,Jonckheere2023May,Schiller2023Nov,Iyer2024May,Thamm2024Apr}. The dominating interpretation of the collider experiments relies on the so-called ``time-domain braiding'' picture~\cite{Han2016Mar,Lee2019Jul}, where an exchange process involving impinging anyons and those spontaneously excited at the QPC provides a dominant, braiding contribution to the noise.

The anyon collider setup is similar to two-particle interferometry of Hong-Ou-Mandel (HOM) type, but with one important difference: In the anyon collider setup, the injected beams of particles are the result of random Poisson processes. In contrast, a HOM interferometer typically uses two controlled, time-delayed injections of particles onto the beam-splitter. This setup was originally implemented for photons (bosons)~\cite{Hong1987Nov}, and was later extended to electrons (fermions) using QH edge states~\cite{Olkhovskaya2008Oct, Bocquillon2012,Jonckheere2012Sep,Bocquillon2013Jan,Dubois2013Aug,Dubois2013Oct,Ferraro2013Nov,Bocquillon2014Jan,Glattli2016Feb,Marguerite2017Mar}, for which inter-edge mode interactions~\cite{Wahl2014Jan,Freulon2015Apr,Marguerite2016Sep,Cabart2018Oct,Rebora2020Jun} and channel mixing~\cite{Taktak2022Oct,Acciai2022Mar,Acciai2023May,Glattli2025Jan} can play a relevant role. A time-controlled anyonic HOM interferometer was only very recently reported~\cite{Ruelle2024Sep}, relying on simulating the injected anyonic states with voltage pulses, as proposed in Ref.~\cite{Jonckheere2023May}.

In HOM interferometry, the quantum statistics of bosons and fermions is manifest as a peak and a dip, respectively, in the so-called HOM ratio (see Eq.~\eqref{eq:HOM_Ratio} below). This quantity is given by the excess tunneling noise produced by two-particle injection, divided by the excess noise from single-particle injections, and quantifies the correlations of emitted particles in the two output channels of the interferometer. For bosons, the noise correlations are peaked around vanishing delay, as the Bose statistics produces a vanishing amplitude for output in two different channels when the particles simultaneously arrive at the beam splitter.
By contrast, fermions produce a dip around vanishing delay, since the Pauli principle prevents two fermions from exiting in the same output state~\cite{Blanter2000Sep,Bocquillon2013Jan}. As (Abelian) anyons are in some sense intermediate between fermions and bosons, it is an interesting, open question how anyonic statistics might manifest in a HOM interferometer setup: a natural expectation is that HOM interferometry of anyons produces features intermediate between those observed for bosons and fermions, like a reduced dip in the HOM ratio. Indeed, a heuristic estimate of the probability of two anyons
exiting a two-particle interferometer in different channels, due to different possible windings of the anyons around each other, can be associated with a statistical factor $(1-\cos\vartheta)/2$~\cite{Campagnano2012Sep}, which is intermediate between the bosonic $(\vartheta=0)$ and fermionic $(\vartheta=\pi)$ scenarios. This factor, however, is based on braiding of anyons in real space, which does not occur in the simple point-like QPC geometry as usually considered in anyon colliders.

\begin{figure}[t!]
    \centering
\includegraphics[width=\columnwidth]{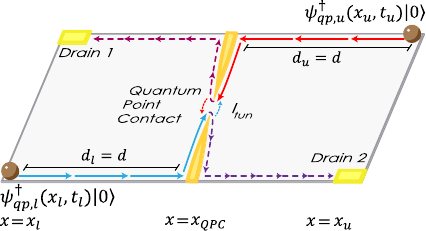}   
    \caption{Hong-Ou-Mandel interferometer realized in a fractional quantum Hall device setup. Point-like anyonic states depicted by brown balls are injected at positions $\xu$, $\xl$ in the upper (u) and lower (l) edges at times $\tu$ and $\tl$, respectively. Drain terminals are used to detect the excess electronic noise $S$ due to anyon interference at the collider quantum point contact at $x=x_\mathrm{QPC}$. The distances $d_\mathrm{u}$ and $d_\mathrm{l}$ are for convenience both taken as $d$.}
    \label{fig:setup}
\end{figure}

Here, we establish in detail that
the expectation that
the anyonic HOM ratio, realized with FQH edge modes in a standard QPC geometry,
has ``intermediate'' features between bosons and fermions, is in fact incorrect. To this end, we consider the setup in Fig.~\ref{fig:setup}, describing the injection of time-delayed, point-like (i.e., with negligible time width) anyonic excitations in a QPC device in the FQH regime at filling $\nu=1/m$ (with $m$ an odd, positive integer).
Such states have been shown to be experimentally relevant, as they can be simulated with ultra-short voltage pulses~\cite{Jonckheere2023May,Ruelle2024Sep}.
Our main finding is that for point-like anyons, the HOM ratio \textit{does not} contain any information about the exchange phase of the injected anyons due to two effects: For small time delay, the total accumulated exchange phase, acquired from time-domain braiding between injected and thermally excited anyons at the QPC, is erased due to two competing sub-processes. For large time-delay, the injected anyons braid instead independently with the thermally activated anyons, but this contribution cancels in the HOM ratio. 

%%%%%%%%%%%%%%%%%%%%%%%%%%%%%%%%%%%%%%%%%%%%%
%%%%%%%%%%%%%%%%%%%%%%%%%%%%%%%%%%%%%%%%%%%%%
%%%%%%%%%%%%%%%%%%%%%%%%%%%%%%%%%%%%%%%%%%%%%%%%%%%%%%%%%%%%%%%%
\textcolor{blue}
{\noindent{\textit{Setup and model.---}}
} We consider the unperturbed, bosonized Hamiltonian~\cite{Wen1995Oct} (we set $\hbar=1$)
\begin{align}
    H_0 = \frac{v_F}{4\pi\nu}\int dx \left[\left(\partial_x\fu(x)\right)^2 + \left(\partial_x\fl(x)\right)^2\right],
\end{align}
where $\fu$ and $\fl$ are bosonic modes propagating to the left and to the right on the upper (u) and lower (l) edge (see Fig.~\ref{fig:setup}), respectively. Both modes propagate with the velocity $v_F$ and obey the commutation relations
\begin{align}\label{eq:commutation_relations}
    \left[\fj(x),\phi_k(y)\right] = \mp i\pi\nu \delta_{jk} \text{sgn}(x-y), \quad j,k=\mathrm{u,l},
\end{align}
where $\delta_{jk}$ is the Kroenecker delta and $\text{sgn}(x)$ is the sign function. Fractionalized quasiparticle excitations on the edges $j=\mathrm{u,l}$ are described by the vertex operators
\begin{align} \label{eq:quasiparticle_operators}
    \psj(x) = \frac{F_j}{\sqrt{2\pi \alpha}} e^{-i\fj(x)}\,,
\end{align}
where $\alpha$ is a short-distance cut-off and $F_j$ are the so-called Klein-factors, obeying the algebra $F_j^{} F_j^\dagger = F_j^{\dagger} F_j^{}=1$ and $F_i^{} F_j^\dagger =-F_j^{\dagger} F_i^{}$ for $i\neq j$. In this work, however, Klein factors always appear in such a way that their product evaluates to unity and we thus ignore them in the following. 

The charge densities propagating on the edges are given by
\begin{align} \label{eq:charge_density}
    \rj(x) \equiv \mp q\frac{\partial_x \fj(x)}{2\pi},
\end{align}
with $-$ for $j=\rm{l}$ and $+$ for $j=\rm{u}$ and where $q$ is the electron charge. Equations ~\eqref{eq:commutation_relations}-\eqref{eq:charge_density} imply the following commutation relation between charge density and vertex operators
\begin{align} 
    \left[\rj(x),\psj^{\dagger}(y)\right] = q\nu\delta(x-y)\psj^{\dagger}(y)\,.
\end{align}
Furthermore, combining Eqs.~\eqref{eq:commutation_relations}-\eqref{eq:quasiparticle_operators} one finds that exchanging the vertex operators $\psj$ at different spatial coordinates results in an exchange phase factor according to
\begin{align} 
\label{eq:anyon_statistics}
    \psj(x)\psj(y) = \psj(y)\psj(x)e^{\pm i\vartheta \text{sgn}(x-y)},
\end{align}
with $+(-)$ for $j=\rm{u}\,(\rm{l})$. %Here, we have fixed $x>y$, so that for $x<y$, $\vartheta\to -\vartheta$.
Moreover, vertex operators~\eqref{eq:quasiparticle_operators} with different $j$ anticommute. We thus see that the operators~\eqref{eq:quasiparticle_operators} describe anyonic excitations with fractional charge $q\nu$ and, upon the identification $\vartheta = \pi\nu$, fractional statistics. These features stand in contrast to the electron excitations, i.e., excitations with charge $q$, which can be written in an analogous way to the fractional excitations of Eq.~(\ref{eq:quasiparticle_operators}), namely
\begin{align} \label{eq:electron_operators}
    \pselj(x) = \frac{F_j}{\sqrt{2\pi \alpha}} e^{-i\fj(x)/\nu}.
\end{align}
These operators can be checked to have the appropriate fermionic exchange phase $\vartheta =\pi$.

Next, to describe tunneling of quasiparticles at the quantum point contact (QPC) at $x=x_{\rm QPC}$, see Fig.~\ref{fig:setup}, we add to $H_0$ the tunneling Hamiltonian
\begin{align}\label{eq:Tunneling_Hamiltonian}
    &H_\mathrm{tun} = \Lambda A + \Lambda^*A^{\dagger},\notag \\
    &A= \psu^{\dagger}(x_{\rm QPC})\psl^{}(x_{\rm QPC}).
\end{align}
Here, $|\Lambda| \ll 1 $ is the weak tunneling amplitude, assumed to be energy-independent, and $A$ is a tunneling operator transferring quasiparticles between the two edges. Importantly, $A$ can be interpreted as creating a quasiparticle-quasihole pair at the QPC. At finite temperature, these pairs are associated with thermal fluctuations; at zero temperature, they are associated with quantum fluctuations. 
%%%%%%%%%%%%%%%%%%%%%%%%%%%%%%%%%%%%%%%%%%%%%%%%%%%%%%%%%%%%%%%%%%%%%%%%%%%%%%
A fundamental ingredient of the two-particle interferometer setup is a tunable injection of single anyonic excitations. In this work, we consider point-like anyon injections, defined in terms of the state
\begin{align}
\label{eq:auxiliary_state}
\ket{\varphi} \equiv \psu^\dagger(x_\mathrm{u},\tu)\psl^{\dagger}(x_\mathrm{l},\tl)\ket{0},      
\end{align}
with $\psj^\dagger(x)$ given in Eq.~\eqref{eq:quasiparticle_operators} and where $\ket{0}\equiv \ket{0}_{\rm u}\otimes\ket{0}_{\rm l}$ is the joint unperturbed, equilibrium ground state of the edges. The state~\eqref{eq:auxiliary_state} describes the injection of two point-like anyons \eqref{eq:quasiparticle_operators} at the upper and lower edge locations $x_{\rm u}$ and $x_{\rm l}$, at times $t_{\rm u}$ and $t_{\rm l}$, respectively. 

It was recently pointed out in Ref.~\cite{Jonckheere2023May} that the expressions for the tunneling currents and noise produced in the QPC by point-like anyon injections are fully equivalent to the combined application of two specially tailored voltage pulses $V_j(t) = \frac{2\pi}{q}\delta(t-t_j)$.
As long as the edge states have a linear dispersion, the injected voltage profiles do not disperse. In this way, the point-like injection considered in this work can be experimentally \textit{simulated} with voltage pulse injections, as previously studied in the context of QH edge states~\cite{Dubois2013Aug,Moskalets2015May,Rech2017Feb,Vannucci2017Jun,Ronetti2018Aug,Ferraro2018Dec,Ronetti2018Aug,Ronetti2019May}.

\textcolor{blue}{\noindent{\textit{HOM noise and ratio.---}}}To analyze noise correlations and exchange statistics of anyons in the HOM setup, we next use the Hamiltonian $H_0+H_{\rm tun}$ and perturbatively compute the tunneling current and noise.

%%%%%%%%%%%%%%%%%%%%%%%%%%%%%%%%%%%%%%%%%%%%%%%%%%%%%%%%%%%%%%%%%%%%%%%%%%%%%%
To leading order in $|\Lambda|$, the tunneling current operator
is given as~\cite{Martin2005Jan_2}
\begin{align}
    &\Itun(t) = iq\nu\left[\Lambda A(t) - \Lambda^* A^{\dagger}(t)\right], \label{eq:tunneling_current_operator}
\end{align}
where
$A(t)$ is the time evolution in the interaction picture.
In this work, operator expectation values are evaluated with respect to the auxiliary state~\eqref{eq:auxiliary_state}, and will be denoted by $\langle \bullet \rangle_\varphi$. As will be clear below, however, such expectation values can be related to those with respect to the equilibrium state, $\langle \bullet \rangle_0$. We have 
\begin{align} \label{eq:Green's_function}
    &\langle A(t) A^{\dagger}(t') \rangle_0 =  
    \langle A^{\dagger}(t) A(t')  \rangle_0 \notag \\
&=\frac{1}{(2\pi \alpha)^2}\left[\frac{\pi k_BT\alpha/v_F}{i\sinh(\pi k_BT (t-t'-i\alpha/v_F))}\right]^{4\delta}\notag\\
 &=\frac{1}{(2\pi \alpha)^2}\left[\frac{\pi k_BT\alpha/v_F}{\sinh(\pi k_BT |t-t'|)}\right]^{4\delta}e^{-i2\pi\delta \mathrm{sgn}(t-t')}.
\end{align}
Here, $T$ is the temperature, $k_B$ is the Boltzmann constant, and in the second line we used that the short distance cutoff $\alpha\ll1$. The exponent $\delta$ is the so-called scaling dimension of the quasiparticle vertex operators~\eqref{eq:quasiparticle_operators}, defined from the expression
\begin{align}
{\langle\psi^\dagger_{\text{qp},j}(0,t)\psi^{}_{\text{qp},j}(0,0)\rangle}_0 \sim t^{-2\delta}.
\end{align}
The scaling dimension thus governs the slow, characteristically power-law, decay of the temporal correlations between quasiparticle-quasihole pairs at the QPC. Generically, $\delta$ is a non-universal parameter susceptible to a broad range of edge effects, e.g., interactions, disorder, neutral modes, and 1/f noise \cite{Kane1992Dec,Pryadko2000Apr,Rosenow2002Feb,Papa2004Sep,Ferraro2008Oct,Braggio2012Sep,Kumar2022Jan,Acciai2025Feb}. It is only in the ideal case with no such effects that $\delta$ is directly related to the FQH filling factor as $\nu=2\delta$, as would be found from evaluating correlations functions of the operators~\eqref{eq:quasiparticle_operators} with respect to $H_0$. Since for ideal Laughlin states, one universally has that $\vartheta=\pi\nu$, the absence of non-universal effects further implies that also the scaling dimension and the statistical exchange phase are directly related as $\vartheta=2\pi\delta$. However, this relation cannot be expected to hold true in realistic devices, and discerning the non-universal effects of $2\pi\delta$ from $\vartheta$ is an essential experimental issue in detecting anyonic statistics.  Throughout this work, we will therefore carefully distinguish the parameters $\delta$ and $\vartheta$, and treat them as two independent variables. 

The key quantity of interest in this work is the low-frequency noise~\cite{Blanter2000Sep,Kobayashi2021Sep} due to tunneling at the QPC in the FQH device. This noise is obtained from
the correlation function
of tunneling current operators~\eqref{eq:tunneling_current_operator} according to~\cite{Martin2005Jan_2,normalization}
\begin{align} \label{eq:HOM_noise}
    S &= \int_{-\infty}^{+\infty}
    dt
    \int_{-\infty}^{\infty}dt'~\langle \left\{\Itun(t), \Itun(t')\right\} \rangle_{\varphi} \notag \\
    &= (q\nu|\Lambda|)^2 
    \int_{-\infty}^{+\infty}
    dt
    \int_{-\infty}^{\infty}dt' \left\langle\left\{A(t),A^{\dagger}(t')\right\} + \mathrm{H.c.}\right\rangle_{\varphi}
\end{align}
where $\{X,Y\}=XY+YX$ is the anticommutator.

In the following, we are interested in the impact that simultaneous or close-to-simultaneous injections of quasiparticle excitations onto the QPC has on the noise. This effect is encoded in the so-called HOM noise, $\Shom$, which is measured when anyons are injected on both edges. To isolate the tunneling noise from the background (thermal) noise, it is customary to subtract  from $\Shom$ the background fluctuations $S_{\mathrm{eq}}$, which are found from Eq.~(\ref{eq:HOM_noise}) by taking $\langle \bullet \rangle_\varphi \to \langle \bullet \rangle_0$. This subtraction defines the experimentally relevant \emph{excess} HOM noise $\Delta \Shom = \Shom - S_{\mathrm{eq}}$. Furthermore, to quantify the effect of two injections, the excess HOM noise is normalized with the corresponding excess Hanbury Brown-Twiss (HBT) noise, $\Delta S_{\text{HBT},j} \equiv S_{\text{HBT},j} - S_{\mathrm{eq}}$, for $j=\rm{u},\rm{l}$, which is obtained when anyons are injected on only one edge. The above procedures are jointly captured with the HOM-noise ratio~\cite{Dubois2013Aug,Wahl2014Jan,Bocquillon2014Jan}
\begin{align}
\label{eq:HOM_Ratio}
    \mathcal{R}(\tau_d) \equiv \frac{\Delta\Shom}{\Delta S_{\mathrm{HBT,u}}+\Delta S_{\mathrm{HBT,l}}},
\end{align}
which is a function of the time delay $\tau_d\equiv t_{\rm l}-t_{\rm u}$ between the two anyon injections. The HOM-noise ratio~\eqref{eq:HOM_Ratio} is the key quantity of interest in this work.
%and in the following sections, we compute $\mathcal{R}(\tau_d)$ for the point-like anyon injections~\eqref{eq:auxiliary_state}.

%%%%%%%%%%%%%%%%%%%%%%%%%%%%%%%%%%%%%%%%%%%%%%%%%%%%%%%%%%%%%%%%%%%%%%%%%%%%%%
\textcolor{blue}{\noindent{\textit{HOM interferometry with point-like anyons.---}}}
We now compute the tunneling noise~\eqref{eq:HOM_noise} and then evaluate the HOM ratio~\eqref{eq:HOM_Ratio}.
For the auxiliary state injection~\eqref{eq:auxiliary_state}, the expressions for the tunneling noise involve non-equilibrium correlation functions of the form $\langle \varphi|A(t) A^{\dagger}(t')|\varphi\rangle$. To compute these functions, we use that the state injection~\eqref{eq:auxiliary_state} is fully equivalent to shifting the vertex operators~\eqref{eq:quasiparticle_operators} with an additional creation of solitons in the bosonic modes $\phi_j$. In other words, we perform the shifts~\cite{Rosenow2016Apr,Schiller2023Nov}
\begin{align}
    &\phi_{u/l}(x,t_{u/l}) \notag \\
    &\rightarrow \phi_{u/l}(x,t_{u/l}) + 2\vartheta\Theta[\mp x-v_F(t-t_{u/l})\pm x_{u/l}],
\end{align}
with $\Theta(\bullet)$ the Heaviside function.
The very same phase shifts arise from the voltage pulses discussed above~\cite{Jonckheere2023May}.
At $x=x_\mathrm{QPC}$, the symmetric setup in Fig.~\ref{fig:setup} produces a constant offset $|x_\mathrm{QPC}-x_{\rm u,l}|=d_{\rm u,l}\equiv d$, which we absorb into the injection times $t_{\rm u,l}$. We then express the chiral evolution of the bosonic modes as 
\begin{align}\label{eq:soliton_phase}
    \phi_{\rm u,l}(t_{\rm u,l}) \rightarrow \phi_{\rm u,l}(t_{\rm u,l}) + 2\vartheta\Theta(t_{\rm u,l}-t).
\end{align}
By construction, the anyons injected into the upper and lower edges thus reach the QPC at the times $t=t_{\rm u,l}$, respectively. The shift~\eqref{eq:soliton_phase} produces a phase factor in the non-equilibrium correlation function~\eqref{eq:Green's_function} and we find
\begin{align}\label{eq:non-equilibrium_correlation_function}
    {\langle A(t) A^{\dagger}(t') \rangle}_{\varphi} = 
    {\langle A(t) A^{\dagger}(t') \rangle}_0 e^{i2\vartheta\Phi(t,t')}\,,
\end{align}
with the time-dependent phase component
\begin{align}
\label{eq:Phi_function}
\Phi(t,t') \equiv \Theta(t_{\rm u}-t)-\Theta(t_{l}-t)+\Theta(t_{l}-t') - \Theta(t_{\rm u}-t').
\end{align}
We thus see that the phase factor in Eq.~\eqref{eq:non-equilibrium_correlation_function} manifests a fractional exchange phase $2\vartheta=2\pi\nu$ which can be interpreted as braiding (i.e., a double exchange) between injected anyons and quasiparticle-quasihole pairs generated at the QPC.

By inserting the correlation function~\eqref{eq:non-equilibrium_correlation_function} into the tunneling  noise~\eqref{eq:HOM_noise}, we obtain the excess HOM noise~\cite{supp}
\begin{align}
    &\Delta \Shom = \frac{4(2q\nu|\Lambda|)^2}{(2\pi \alpha)^2}\left(2\pi k_B T\right)^{4\delta-1}\left(\frac{\alpha}{v_F}\right)^{4\delta}\cos{(2\pi\delta)}\notag\\
    &\times \left[\cos{(2\vartheta)}-1\right]\int_{0}^{|\tau_d|}dt~ \mathcal{B}\left(e^{-2\pi t/\beta};2\delta,\gamma\right).
    \label{eq:excess_noise}    
\end{align}
In Eq.~\eqref{eq:excess_noise}, $\mathcal{B}\left(x;a,b\right)$ is the incomplete Beta function, $\beta^{-1} \equiv k_BT$, $\gamma = 1-4\delta$, and $\tau_d \equiv t_{\rm l} - t_{\rm u}$ is the time delay. We see from Eq.~\eqref{eq:excess_noise} that the excess noise vanishes for zero time delay $\tau_d = 0$. This feature follows from Eq.~\eqref{eq:non-equilibrium_correlation_function} since (neglecting the energy dependence in the tunneling amplitude $\Lambda$ and screening effects~\cite{Christen1996Jul,Pretre1996Sep,Meair2013Jan,Dashti2021Dec}), the device is in equilibrium at zero delay:
$\langle A(t) A^{\dagger}(t') \rangle_{\varphi} = \langle A(t) A^{\dagger}(t') \rangle_0$.
Moving on to the HOM ratio~\eqref{eq:HOM_Ratio}, we find that it can be written in compact form as~\cite{supp}
\begin{align}
\label{eq:HOM_ratio_point}
    \mathcal{R}(\tau_d) &=1-\frac{\int_{0}^{\infty} dt~\mathcal{B}\left(e^{-2\pi(t+|\tau_d|)/\beta};2\delta,\gamma\right)}{\int_{0}^{\infty} dt~\mathcal{B}\left(e^{-2\pi t/\beta};2\delta,\gamma\right)}.
\end{align}
%%%%%%%%%%%%%%%%%%%%%%%%%%%%%%%%%
\begin{figure}[t!]
    \centering
\includegraphics[width=\columnwidth]{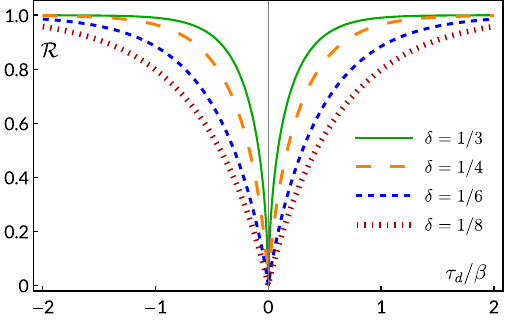}  
    \caption{Dimensionless HOM ratio $\mathcal{R}$ [Eq.~\eqref{eq:HOM_Ratio}] as a function of the injection time delay $\tau_d\equiv t_{\rm l}-t_{\rm u}$ (in units of inverse temperature $\beta$), for several scaling dimensions $\delta$.
    } 
    \label{fig:Anyon_HOM_ratio}
\end{figure}
%%%%%%%%%%%%%%%%%%%%%%%%%%%%%%%%%
Crucially, we see that the anyon exchange phase $\vartheta$ is fully absent in the HOM ratio when point-like anyons are injected. Instead, the HOM ratio strongly depends on the scaling dimension $\delta$ of the quasiparticle-quasihole pairs excited at the QPC. This feature is shown in Fig.~\ref{fig:Anyon_HOM_ratio}, where we plot $\mathcal{R}(\tau_d)$ for different values of $\delta$. Besides the strong dependence on $\delta$, the anyonic HOM ratio has another distinct feature not shared with noninteracting electrons, namely its temperature dependence. It was shown in 
Ref.~\cite{Jonckheere2023May} that the width of the HOM curves increases with decreasing temperature, in contrast to the free-electron case, where the width is set only by the temporal extension of the injected states, without any temperature dependence~\cite{Jonckheere2012Sep,Glattli2016Feb,supp}. At small time delays $\tau_d\ll\beta$, Eq.~\eqref{eq:HOM_ratio_point} simplifies to~\cite{Jonckheere2023May}
\begin{align}
\label{eq:R_simplified}
    &\mathcal{R}(\tau_d)\approx 1-e^{-|\tau_d|/\tau_\mathrm{th}}\,,
\end{align}
where the parameter $\tau_\mathrm{th}\equiv\beta/(4\pi\delta)$ defines a characteristic thermal time scale for anyon correlations, governing the QPC quasiparticle-quasihole pair time correlations. Importantly, we see that this time scale does not involve $\vartheta$ but only $\delta$. It follows that unlike for electrons and bosons, the standard HOM ratio~\eqref{eq:HOM_Ratio} does not probe the exchange statistics angle $\vartheta$ of point-like, injected anyons. We now elucidate why this is the case.

\textcolor{blue}{\noindent{\textit{Exchange-phase erasure.---}}}The absence of the exchange phase $\vartheta$ in the standard HOM ratio~\eqref{eq:HOM_ratio_point} can be understood within the anyon time-domain braiding picture~\cite{Lee2019Jul}. To illustrate this, we first rearrange the correlation function~\eqref{eq:non-equilibrium_correlation_function} as
\begin{align}
\label{eq:rearranged_non-equilibrium_correlation_function}
&\langle A(t) A^{\dagger}(t') \rangle_{\varphi} = \langle A(t) A^{\dagger}(t') \rangle_0\notag \\&\times e^{i2\vartheta\left[\Theta(t_{\rm u}-t) - \Theta(t_{\rm u}-t')\right]}\,e^{i2\vartheta\left[\Theta(t_{\rm l}-t')-\Theta(t_{\rm l}-t)\right]}. 
\end{align}
We also rewrite the tunneling noise~\eqref{eq:HOM_noise} as
\begin{align} 
\label{eq:HOM_interference_noise}
    &S^{}\propto \int_{-\infty}^{\infty} dt \left(\int_{t}^{\infty} + \int_{-\infty}^{t} \right) dt'\sum_{k=\pm} {}_k\langle t,\tau_d|t',\tau_d\rangle_k+ \mathrm{H.c.}
\end{align}
%%%%%%%%%%%%%%%%%%%%%%%%%%%
\begin{figure}[t!]
  \centering
  \includegraphics[width=\columnwidth]{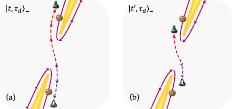}
  \caption{Depiction of a tunneling sub process that contributes to the noise in Eq.~\eqref{eq:HOM_interference_noise}, here with $k=-$ and for $t'>t_{\rm u},t_{\rm l}>t$. Injected anyons are depicted as brown balls with solid, purple trajectories. Anyonic quasiparticle and quasiholes excited at the QPC are depicted by green peaks with dashed, red trajectories and white peaks with dashed violet trajectories, respectively. (a) A quasiparticle-quasihole pair is excited at the QPC \emph{before} the arrival of two injected anyons. (b) A quasiparticle-quasihole pair is excited \emph{after} after the arrival of the injected anyons.}
  \label{fig:injected_anyons-QPC_excitations}
\end{figure}
%%%%%%%%%%%%%%%%%%%%%%%%%%%
Here, $|\bullet,\tau_d\rangle_- \equiv A(\bullet)\ket{\varphi}$ denotes the state with a quasiparticle created on the upper edge and a quasihole on the lower edge. Likewise, $|\bullet,\tau_d\rangle_+ \equiv A^{\dagger}(\bullet)\ket{\varphi}$ has a quasihole on the upper edge and a quasiparticle on the lower edge.
We now begin by examining the contribution to the integral~\eqref{eq:HOM_interference_noise} for $t'>t$. In this range,
%\sout{We now first examine the contribution of the integral in Eq.~\eqref{eq:HOM_interference_noise} within the range $t'\in(t,+\infty)$. Here,}
the states $\ket{t,\tau_d}_{-}$ and $\ket{t',\tau_d}_{-}$, describe quasiparticle-quasihole excitations created \emph{before} (Fig.~\ref{fig:injected_anyons-QPC_excitations}\textcolor{blue}{a}) and \emph{after} (Fig.~\ref{fig:injected_anyons-QPC_excitations}\textcolor{blue}{b}) the arrival of the injected anyons at the QPC, respectively. Furthermore, the conjugated states $_{-}\!\bra{t,\tau_d}$ and $_{-}\!\bra{t',\tau_d}$
%in Eq.~\eqref{eq:HOM_interference_noise}
can be visually represented by rewinding the quasiparticle paths %associated to $\ket{t,\tau_d}_{-}$
as shown in Fig.~\ref{fig:interference_loops}\textcolor{blue}{a-b}. In this way, the inner products of these states can be interpreted as time-domain interference loops where the injected anyons can braid with the QPC quasiparticle-quasihole excitations depending on the time delay $\tau_d$. 

Crucially, the time-domain braiding processes discussed above occur when at least one of the injection times $\tu,\tl$ falls within the window $(t,t')$. Of all possible intervals $(t,t')$, the most relevant ones are those for which $t'-t\lesssim\beta$, the contributions from others being suppressed by the decay of the correlation function. There are therefore three scenarios, as shown in Fig.~\ref{fig:interference_loops}\textcolor{blue}{c}: If the injection-time separation $\tau_d\ll\beta$, both $\tu$ and $\tl$ can fall in the window $(t,t')$ and both injected anyons braid with the locally excited ones. However, the two anyons accumulate opposite braiding phases, resulting in a cancellation. If instead $\tau_d\gg\beta$, only one of the injected anyons can fall within the ``braiding window'', contributing to the noise with a term proportional to $\cos(2\vartheta)$.
   %%%%%%%%%%%%%%%%%%%%%%%%%%%%%%
\begin{figure}[t!]
  \centering
  \includegraphics[width=\columnwidth]{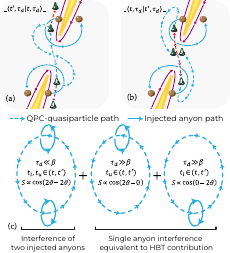}
  \label{fig:interferences_ttprime}
  \caption{Time-domain braiding amplitudes for the processes depicted in Fig.~\ref{fig:injected_anyons-QPC_excitations}.
  (a) The amplitude is composed of Fig.~\ref{fig:injected_anyons-QPC_excitations}\textcolor{blue}{a} with the time-reverse of Fig.~\ref{fig:injected_anyons-QPC_excitations}\textcolor{blue}{b}. (b) The amplitude is composed of Fig.~\ref{fig:injected_anyons-QPC_excitations}\textcolor{blue}{b} with the time-reverse of Fig.~\ref{fig:injected_anyons-QPC_excitations}\textcolor{blue}{a}. The blue solid and dashed lines thus depict \emph{time-reversed} paths of injected anyons and QPC quasiparticle-quasihole excitation, respectively. Trajectories at later times cross \textit{above} trajectories at earlier times. (c) The braiding links to the left correspond to the two processes in (a-b). Exchange-phase effects in these processes for $\tau_d \ll \beta$ are erased from the HOM noise through counter-braiding in the time-domain. The middle and right braiding links are formed when $\tau_d \gg \beta$ for single anyon injection, which are thus HBT contributions.
  }
  \label{fig:interference_loops}
\end{figure}
%%%%%%%%%%%%%%%%%%%%%%%%%%%%
%For $\tau_d\ll \beta$, each injected anyon arrives at the QPC within the ``interference window'', i.e., the $(t',t)$ such that $t'>t_{\rm l},t_{\rm u}>t$.
%Thus, both injected anyons braid in the time-domain with the locally excited anyons, as depicted in Figs.~\ref{fig:interference_loops}\textcolor{blue}{a-b}. However, the two anyons accumulate opposite braiding phases, resulting in a cancellation.
%In the opposite limit, $\tau_d\gg \beta$, processes with $t'>t_{\rm u}>t$ or $t'>t_{\rm l}>t$ are favored, and only one of the injected anyons participates in time-domain braiding, producing terms proportional to $\cos{(2\vartheta)}$.
Only such processes contribute to the appearance of $\vartheta$ in $S_\mathrm{HOM}$~\footnote{However, $\vartheta$ appears in the noise jointly with the scaling dimension $\delta$, making an extraction of only $\vartheta$ challenging.}. 
Moreover, at large time delays $\tau_d\gg\beta$, there are independent anyon injections on the two edges, which then corresponds to the HBT configuration.
%Moreover, by definition, these contributions correspond to the HBT noise $S_{\rm HBT}$.
The same phase information is therefore contained in the numerator and denominator of the HOM ratio $\mathcal{R}(\tau_d)$, leading to the absence of $\vartheta$ in the HOM ratio, and resulting in Eq.~\eqref{eq:HOM_ratio_point}.
A perfectly analogous analysis holds for the second contribution in
%the time integral with $t'\in(-\infty,t)$ in
Eq.~\eqref{eq:HOM_interference_noise}, namely, the time integral with $t'<t$. Figure~\ref{fig:interference_loops}\textcolor{blue}{c} depicts the time-domain braiding sub-processes at the QPC for $k=-$. The analogous processes for $k=+$ are not drawn, but they are readily obtained by reversing the charge of the excited quasiparticles at the QPC.

According to the above analysis, we conclude that the counterbalancing braiding sub-processes, promoted by the anyonic long-time correlations
do not allow direct observation of the exchange statistics based on the standard HOM ratio. This is the main result of this work. We further remark that our derivation of the HOM ratio is not straightforward to adapt to non-interacting electrons by setting $\nu=1$ and $\delta=1/2$. As pointed out in Ref.~\cite{Schiller2023Nov}, a bosonic shift~\eqref{eq:soliton_phase} for $\nu=1$, is equivalent to no shift at all. To obtain the HOM ratio in this case, one can use a bosonic shift with a small but non-vanishing temporal width~\cite{supp}.
In this way, the braiding becomes trivial ($2\vartheta=2\pi$) and one recovers that $\mathcal{R}$ gives the wavefunction overlap of the two injected single-electron states~\cite{Jonckheere2012Sep}.

%%%%%%%%%%%%%%%%%%%%%%%%%%%%%%%%%%%%%%%%%%%%%%%%%%%%%%%%%%%%%%%%%%%%%%%%%%%%%%
\textcolor{blue}{\noindent{\textit{Summary and outlook.---}}}We studied two-particle interferometry of point-like anyons in a fractional quantum Hall realization of the Hong-Ou-Mandel (HOM) interferometer setup. We found that, in contrast to bosons~\cite{Hong1987Nov} and fermions~\cite{Bocquillon2013Jan}, the characteristic HOM ratio $\mathcal{R}(\tau_d)$ [see Eq.~\eqref{eq:HOM_Ratio}], with $\tau_d$ the injection-time delay, is void of the anyonic exchange phase $\vartheta$. This result is essentially due to the peculiar nature of the FQH interferometer, that, to leading order in the tunneling amplitude, probes the time-domain braiding of the injected anyons with those excited at the QPC, rather then direct ``collisions'' of the incoming anyons~\cite{Lee2019Jul}. More specifically, our detailed analysis showed that the origins of this absence lie in two complementary effects: When $\tau_d$ is small in comparison to the thermal timescale, the exchange phase accumulated from time-domain braiding between injected anyons and QPC quasiparticle-quasihole pairs is erased due to two sub-processes, whose phase contributions cancel exactly. Instead, for $\tau_d$ large compared to the thermal timescale, there are
processes retaining the exchange phase, but their contributions are canceled in the HOM ratio
$\mathcal{R}(\tau_d)$. This feature shows a clear difference between anyonic HOM interferometers and their fermionic/bosonic counterparts.
Hence, extracting the anyonic phase from HOM interferometry requires going beyond the standard HOM ratio and implement more sophisticated measurement protocols. One such approach was recently investigated in Ref.~\cite{Ruelle2024Sep}, where estimates of both scaling dimension and the anyonic exchange phase were provided from combined measurements of the tunneling conductance and noise.

As a natural follow up of our work, it would be interesting to investigate whether a full exchange-phase cancellation in the HOM ratio persists for finite-width anyon states~\cite{Schiller2023Nov,Thamm2024Apr,Iyer2024May}. Additionally, it remains an open question whether a wave-function description of anyonic Levitons~\cite{Sim} can be related to the weak backscattering HOM noise considered in this work.
Another natural direction is to investigate whether there are possibilities to extract the statistics of anyons on  more complex edges, e.g., at fillings $\nu=2/5$ or $\nu=2/3$, whose description requires taking into account non-topological edge effects, e.g., interactions, disorder, and equilibration~\cite{Kane1994Jun,Protopopov2017,Nosiglia2018,Spanslatt2021Sep,Hein2023Jun}.

\textit{Note:} Some of the results in this paper have been reported in the Master thesis in Ref.~\cite{Varada2023}.
%%%%%%%%%%%%%%%%%%%%%%%%%%%%%%%%%%%%%%%%%%%%%%%%%%%%%%%%%%%%%%%%%%%%%%%%%%%%%%
\textcolor{blue}{\noindent{\textit{Acknowledgments.---}}}
We gratefully acknowledge Janine Splettstoesser for discussions and feedback on the manuscript. This project has received funding from the European Union’s Horizon 2020 research and innovation programme under grant agreement No.~862683 (FET-OPEN UltraFastNano) and No. 101031655 (TEAPOT). We also acknowledge funding from the Swedish Vetenskapsr\r{a}det via Project No.~2023-04043 (C.S.), from the Area of Advance Nano at Chalmers University of Technology (C.S), and from the PNRR MUR Project No.~PE0000023-NQSTI (M.A.). \\
%%%%%%%%%%%%%%%%%%%%%%%%%%%%%%%%%%%%%%%%%%%%%
\noindent{\textit{Data availability.---}}
Data supporting the findings of this article are openly available~\cite{Varada2025May}.

\clearpage
\begin{onecolumngrid}
%%%%%%%%%%%%%%%%%%%%%%%%%%%%%%%%%%%%%%%%%%%%%%%%%%%%%%%%%%%%%%%%%%%%%%%%%%%%%%
\global\long\def\thesection{S\Alph{section}}
\global\long\def\thesubsection{\Roman{subsection}}
\setcounter{equation}{0}
\setcounter{figure}{0}
\setcounter{table}{0}
\setcounter{page}{1}
\renewcommand{\theequation}{S\arabic{equation}}
\renewcommand{\thefigure}{S\arabic{figure}}

\begin{center}
\large{\bf Supplemental Material for ``Exchange-phase erasure in anyonic Hong-Ou-Mandel interferometry''}
\end{center}
\begin{center}
Sushanth Varada$^{1,2}$,
Christian Sp\r{a}nsl\"{a}tt$^{3,1}$, and Matteo Acciai$^{4,1}$\\
{\it $^{1}$Department of Microtechnology and Nanoscience (MC2),\\Chalmers University of Technology, S-412 96 G\"oteborg, Sweden\\
$^{2}$Department of Physics and Astronomy, Uppsala University, Box 516, S-751 20 Uppsala, Sweden\\
$^{3}$Department of Engineering and Physics, Karlstad University, Karlstad, Sweden\\
$^4$Scuola Internazionale Superiore di Studi Avanzati (SISSA), Via Bonomea 365, 34136 Trieste, Italy}\\
\end{center}
This Supplemental Material contains two sections. In Sec.~\ref{app:calculations_point_anyons}, we provide details of the derivation of the anyonic HOM ratio~(\textcolor{blue}{14})
%\eqref{eq:HOM_Ratio}
in the main text. For useful comparison, we provide in Sec.~\ref{app:electrons} also the analogous calculation of the HOM ratio for non-interacting electrons.

\section{Details of anyon HOM ratio derivation}
\label{app:calculations_point_anyons}
Here, we present the derivation of the HOM ratio~\eqref{eq:HOM_Ratio}
in the main text. For completeness, we also compute the average tunneling current, given as
\begin{align}
    \left\langle \Itun(t)\right\rangle = q\nu|\Lambda|^2\int_{-\infty}^{t}dt'\left\langle\left[A(t),A^{\dagger}(t')\right] - \mathrm{H.c.}\right\rangle_{\varphi}. \label{eq:exp_tunneling_current_operator}
\end{align}
and then the excess HOM noise~\eqref{eq:excess_noise}
To this end, we first insert the correlation function~\eqref{eq:non-equilibrium_correlation_function}
into the current in Eq.~\eqref{eq:exp_tunneling_current_operator} and obtain 
\begin{align}
&\begin{aligned}
    \langle \Itun(t)\rangle &= \frac{q\nu|\Lambda|^2}{(2\pi\alpha)^2}\int_{-\infty}^{t}dt'~ \left[\frac{\pi k_BT\alpha/v_F}{\sinh(\pi k_BT |t-t'|)}\right]^{4\delta}\left(e^{-i2\pi\delta \text{sgn}(t-t')} - e^{-i2\pi\delta \text{sgn}(t'-t)}\right)\left(e^{i2\vartheta\Phi(t,t')} - e^{-i2\vartheta\Phi(t,t')}\right)\\
    &= \frac{4q\nu|\Lambda|^2}{{(2\pi\alpha)^2}}\sin{(2\pi\delta)}\int_{-\infty}^{t}dt'~\left[\frac{\pi k_BT\alpha/v_F}{\sinh(\pi k_BT (t-t'))}\right]^{4\delta}\sin{(2\vartheta\Phi(t,t'))}\,.
\end{aligned}&
\end{align}
To proceed, we focus on the sine function with the time-dependent quantity $\Phi(t,t')$, defined in Eq.~\eqref{eq:Phi_function}
%(\textcolor{blue}{18})
in the main text, and simplify the time integrals. Assuming  $t_{\rm l}>t_{\rm u}$, we next examine the conditions imposed by the arrival times $t_{\rm u,l}$ and the temporal parameter $t$:

For $t>t_{\rm l}>t_{\rm u}$, the Heaviside functions $\Theta(t_{\rm l}-t)$ and $\Theta(t_{\rm u}-t)$ in $\Phi(t,t')$ vanish and the tunneling current is non-zero only when the argument $t’$ falls within the arrival-time window $(t_{\rm l},t_{\rm u})$. We then have
\begin{align}
   \int_{-\infty}^{t}dt' \sin{\{2\vartheta\left[-\Theta(t_{\rm l}-t)+\Theta(t_{\rm u}-t)+\Theta(t_{\rm l}-t') - \Theta(t_{\rm u}-t')\right]\}} = \int_{t_{\rm u}}^{t_{\rm l}}dt' \sin{(2\vartheta)},\quad t>t_{\rm l}>t_{\rm u}.
\end{align}
Similarly, for $t_{\rm l}>t>t_{\rm u}$ the nonzero function $\Theta(t_{\rm u}-t')$ gives a non-trivial tunneling current for $t_{\rm u}>t’$ as
\begin{align}
   \int_{-\infty}^{t}dt' \sin{\{2\vartheta\left[-\Theta(t_{\rm l}-t)+\Theta(t_{\rm u}-t)+\Theta(t_{\rm l}-t') - \Theta(t_{\rm u}-t')\right]\}} = -\int_{-\infty}^{t_{\rm u}}dt' \sin{(2\vartheta)}, \quad t_{\rm l}>t>t_{\rm u}.
\end{align}
Finally, the case $t_{\rm l}>t_{\rm u}>t$ produces only equilibrium conditions and thus a vanishing tunneling current. Combining all three cases above, we arrive at the following expression for the tunneling current
\begin{align}
    &\langle \Itun(t) \rangle_\varphi =\frac{4q\nu|\Lambda|^2}{(2\pi\alpha)^2}\sin{(2\pi\delta)}\sin{(2\vartheta)}\Theta(t-t_{\rm u})\left\{\Theta(t-t_{\rm l})\int_{-\infty}^{t_{\rm l}}dt'\left[\frac{\pi k_BT\alpha/v_F}{\sinh(\pi k_BT (t-t'))}\right]^{4\delta}-\int_{-\infty}^{t_{\rm u}}dt'\left[\frac{\pi k_BT\alpha/v_F}{\sinh(\pi k_BT (t-t'))}\right]^{4\delta}\right\}. 
\end{align}

For $t_{\rm u} > t_{\rm l}$, we see that the arrival times $t_{\rm u,l}$ are simply exchanged in this expression. By next introducing the time delay $\tau_d \equiv t_{\rm l} - t_{\rm u}$ and some minor algebra, we obtain for the tunneling current: 
\begin{align}\label{appendix_eq:current}
    \langle \Itun(t)\rangle_\varphi = \frac{4q\nu|\Lambda|^2}{(2\pi\alpha)^2}\sin{(2\vartheta)}\sin{(2\pi\delta)}\Theta(t)\left\{\Theta\left(t-|\tau_d|\right)\int_{-\infty}^{|\tau_d|}dt'\left[\frac{\pi k_BT\alpha/v_F}{\sinh(\pi k_BT (t-t'))}\right]^{4\delta}-\int_{-\infty}^{0}dt'\left[\frac{\pi k_BT\alpha/v_F}{\sinh(\pi k_BT (t-t'))}\right]^{4\delta}\right\}.
\end{align}
Next, we rewrite the integrals over $t'$ in terms of the incomplete Beta function
\begin{align}
    \mathcal{B}\left(x;a,b\right) \equiv \int_0^x y^{a-1}(1-y)^{b-1}.
\end{align} 
For generic  $z$, we then have
\begin{align}
    \int_{-\infty}^{z}dt' \left[\frac{\pi k_B T \alpha/v_F}{\sinh(\pi k_B T (t-t'))}\right]^{4\delta} = \left(2\pi k_B T \frac{\alpha}{v_F}\right)^{4\delta} \int_{-\infty}^{z}dt'~ \left[\frac{1}{e^{\pi k_B T (t-t')} - e^{-\pi k_B T (t-t')}}\right]^{4\delta}
\end{align}
and upon the charge of variable $y = e^{-2\pi k_B T (t-t')}$, we obtain the formula
\begin{align}\label{appendix_eq:Beta_function}
    (2\pi k_B T)^{4\delta-1} \left(\frac{\alpha}{v_F}\right)^{4\delta} \int_{0}^{e^{-2\pi k_{B} T (t-z)}} dy~(1-y)^{(1-4\delta)-1}y^{2\delta-1} = (2\pi k_B T)^{4\delta-1} \left(\frac{\alpha}{v_F}\right)^{4\delta}~\mathcal{B}(e^{-2\pi k_B T (t-z)},2\delta, 1-4\delta).
\end{align}
Using this result in Eq.~\eqref{appendix_eq:current}, we obtain the result
\begin{align}
    \langle \Itun(t)\rangle_{\varphi} = \frac{4q\nu|\Lambda|^2}{(2\pi\alpha)^2}(2\pi k_B T)^{4\delta-1} \left(\frac{\alpha}{v_F}\right)^{4\delta}\sin{(2\vartheta)}\sin{(2\pi\delta)}\Theta(t)\Big[\Theta\left(t-|\tau_d|\right)\mathcal{B}\left(e^{-2\pi(t-|\tau_d|)/\beta};2\delta,\gamma\right)-\mathcal{B}\left(e^{-2\pi t/\beta};2\delta,\gamma\right) \Big],\label{eq:tunneling_current}
\end{align}
in agreement with Ref.~\cite{Schiller2023Nov}, apart from having $2\delta$ instead of $1+2\delta$ as the second argument of the Beta function.

Moving on to the tunneling noise, by inserting the correlation function \eqref{eq:non-equilibrium_correlation_function}
%(\textcolor{blue}{17})
into the symmetrized zero-frequency noise~\eqref{eq:HOM_noise}
%(\textcolor{blue}{13})
in the main text, we express the excess HOM noise as 
\begin{align}
    \Delta S_{\rm HOM} &= \frac{(2q\nu|\Lambda|)^2}{(2\pi\alpha^2)}\cos{(2\pi\delta)} \notag \\
    &\times \int_{-\infty}^{\infty}dt\left(\int_{t}^{\infty}dt'~\left[\frac{\pi k_BT\alpha/v_F}{\sinh(\pi k_BT (t'-t))}\right]^{4\delta} + \int_{-\infty}^{t}dt'~\left[\frac{\pi k_BT\alpha/v_F}{\sinh(\pi k_BT (t-t'))}\right]^{4\delta}\right)\left[\cos{(2\vartheta\Phi(t,t'))}-1\right]. 
\end{align}
To proceed, we first consider the integral over $t'$ from $t$ to $+\infty$ and investigate the cosine term in integrand term, assuming $t_{\rm l}>t_{\rm u}$. We note that this integral has limits opposite to those in the tunneling current. As such, the condition $t>t_{\rm l}>t_{\rm u}$ corresponds here to the equilibrium state leading to a vanishing excess-noise contribution. For $t_{\rm l}>t>t_{\rm u}$, we instead obtain a finite noise contribution only when $t_{\rm l}<t'$:
\begin{align}
   \int_{t}^{\infty}dt'\{\cos{(2\vartheta[-1 + \Theta(t_{\rm l}-t')])}-1\} = \int_{t_{\rm l}}^{\infty}dt'~[\cos{(2\vartheta)}-1]\, \quad t_{\rm l}<t'.
\end{align}
In the same way, for $t_{\rm l}>t_{\rm u}>t$, the Heaviside functions depending on $t$ cancel out each other and the non-zero functions contribute to the noise only when $t_{\rm l}>t’>t_{\rm u}$. Then 
\begin{align}
   \int_{t}^{\infty}dt'~\{\cos{(2\vartheta[ \Theta(t_{\rm l}-t')-\Theta(t_{\rm u}-t')])}-1\} = \int_{t_{\rm u}}^{t_{\rm l}}dt'~[\cos{(2\vartheta)}-1], \quad t_{\rm l}>t’>t_{\rm u}.
\end{align}
The second integral over $t'$ from $-\infty$ to $t$ has an equal contribution, and combining these results gives
\begin{align}\label{appendix_eq:excess_HOM_noise}
    \Delta S_{\rm HOM} = 2\frac{(2q\nu|\Lambda|)^2}{(2\pi\alpha)^2}\cos{(2\pi\delta)}[\cos{(2\vartheta)}-1]\int_{t_{\rm u}}^{t_{\rm l}}dt~\left\{\int_{t_{\rm l}}^{\infty}dt'\left[\frac{\pi k_BT\alpha/v_F}{\sinh(\pi k_BT (t'-t))}\right]^{4\delta} + \int_{-\infty}^{t_{\rm u}}dt'\left[\frac{\pi k_BT\alpha/v_F}{\sinh(\pi k_BT (t-t'))}\right]^{4\delta}\right\}.
\end{align}
Analogous to the tunneling current calculations, accounting for the case $t_{\rm u}>t_{\rm l}$ and plugging in the tunable time delay $\tau_d$, gives after some minor algebra and a change of variables 
\begin{align}\label{appendix_eq:excess_HOM_noise2}
    \Delta S_{\rm HOM} = 4\frac{(2q\nu|\Lambda|)^2}{(2\pi\alpha)^2}\cos{(2\pi\delta)}[\cos{(2\vartheta)}-1]\int_{0}^{|\tau_d|}dt\int_{0}^{\infty}dt'\left[\frac{\pi k_BT\alpha/v_F}{\sinh(\pi k_BT (t-t'))}\right]^{4\delta}.
\end{align}

Then, using Eq.~\eqref{appendix_eq:Beta_function} in Eq.~\eqref{appendix_eq:excess_HOM_noise2} leads to the expression~\eqref{eq:excess_noise}
%(\textcolor{blue}{19})
presented in the main text. 

Furthermore, using the HBT auxiliary state $\ket{\varphi} \equiv \psi_{\text{qp},j}^\dagger(x_j,\tj)\ket{0}$, for $j=\rm{u},\rm{l}$, and repeating the above calculations starting from Eq.~\eqref{eq:HOM_noise}
%(\textcolor{blue}{13})
with the time-dependent phase components $\Phi_{\text{HBT},j}(t,t') = \left[\Theta(t_j-t) - \Theta(t_j-t')\right]$ gives the excess HBT noise as
\begin{align}
    \Delta S_{\text{HBT},j} = 2\frac{(2q\nu|\Lambda|)^2}{(2\pi\alpha)^2}\cos{(2\pi\delta)}\left[\cos{(2\vartheta)}-1\right](2\pi k_B T)^{4\delta-1} \left(\frac{\alpha}{v_F}\right)^{4\delta}\int_{0}^{\infty}dt~ \mathcal{B}\left(e^{-2\pi k_BT t};2\delta,\gamma\right).
\end{align}
Accordingly, the standard HOM ratio \eqref{eq:HOM_Ratio}
%(\textcolor{blue}{14})
becomes
\begin{align}
    \mathcal{R}(\tau_d) = \frac{\int_{0}^{|\tau_d|} dt~\mathcal{B}\left(e^{-2\pi k_BT t};2\delta,\gamma\right)}{\int_{0}^{\infty} dt~\mathcal{B}\left(e^{-2\pi k_BT t};2\delta,\gamma\right)} = 1-\frac{\int_{0}^{\infty} dt~\mathcal{B}\left(e^{-2\pi(t+|\tau_d|)/\beta};2\delta,\gamma\right)}{\int_{0}^{\infty} dt~\mathcal{B}\left(e^{-2\pi t/\beta};2\delta,\gamma\right)}.
\end{align}
By further re-writing the numerator of this ratio on an exponential form, we finally arrive at
\begin{align}
&\begin{aligned}
    \mathcal{R}(\tau_d) &= 1 - \frac{1}{\int_{0}^{\infty}dt~\mathcal{B}\left(e^{-2\pi t/\beta};2\delta,\gamma\right)}\int_{0}^{\infty}dt\int_{0}^{\infty}dt' \left[e^{\pi (t+t')/\beta}e^{\pi |\tau_d|/\beta} - e^{-\pi (t+t')/\beta}e^{-\pi |\tau_d|/\beta}\right]^{-4\delta},\\
    & = 1 - \frac{e^{-4\pi\delta |\tau_d|/\beta}}{\int_{0}^{\infty}dt~\mathcal{B}\left(e^{-2\pi t/\beta};2\delta,\gamma\right)}\int_{0}^{\infty}dt\int_{0}^{\infty}dt' \left[e^{\pi (t+t')/\beta} - e^{-[\pi (t+t')+2\pi |\tau_d|]/\beta}\right]^{-4\delta}.
\end{aligned}&
\end{align}
In the regime $2\pi|\tau_d|/\beta \ll 1$, the integral in the numerator simplifies and we obtain
\begin{align}
    \mathcal{R}(\tau_d) \approx 1 - e^{-4\pi\delta|\tau_d|/\beta}
    %= 1 - e^{-2\pi(2\delta)|\tau_d|/\beta} \equiv 1 - e^{-|\tau_d|/\tau_{\delta}}
    ,
\end{align}
which is Eq.~\eqref{eq:R_simplified}
%(\textcolor{blue}{21})
in the main text, and in agreement with Ref.~\cite{Jonckheere2023May}. 
%%%%%%%%%%%%%%%%%%%%%%%%%%%%%%%%%%%%%%%%%%%%%%%%%%%%%%%%%%%%%%%%%%%%%%%%%%%%%%

\section{HOM ratio for electrons}
\label{app:electrons}
For instructive purposes, we provide here a derivation of the HOM ratio for non-interacting electrons. As outlined in the main text, the approach in Appendix~\ref{app:calculations_point_anyons} cannot be adapted straightforwardly to noninteracting electrons by simply setting $\nu=1$, $\delta=1/2$, and $\vartheta = \pi$. Instead, one can use a bosonic shift with finite temporal width $w>0$~\cite{Schiller2023Nov}, which modifies the soliton profiles of the bosonic modes \eqref{eq:soliton_phase}
%(\textcolor{blue}{16})
as
\begin{align}
\label{eq:shift_with_width}
    \phi_{\rm u,l}(t_{\rm u,l}) \rightarrow \phi_{\rm u,l}(t_{\rm u,l}) + 2\vartheta\left[\frac{1}{\pi}\arctan\left(\frac{t_{\rm u,l}-t}{w}\right)+\frac{1}{2}\right].   
\end{align}
As the width $w\to 0$,  Eq.~\eqref{eq:shift_with_width} reduces to Eq.~\eqref{eq:soliton_phase}
%(\textcolor{blue}{16})
with $\vartheta=\pi$, which amounts to a bosonic shift of $2\pi=0\, (\text{mod}\, 2\pi$), i.e., no shift at all. The transformation~\eqref{eq:shift_with_width} alters the time-dependent phase component $\Phi$ in the correlation function \eqref{eq:non-equilibrium_correlation_function}
%(\textcolor{blue}{17})
as
\begin{align}
\Phi(t,t') = \frac{2}{\pi}\left[\arctan(\frac{t_{\rm u}-t}{w}) - \arctan(\frac{t_{\rm l}-t}{w}) +\arctan(\frac{t_{\rm l}-t'}{w}) - \arctan(\frac{t_{\rm u}-t'}{w})\right].
\end{align}
By inserting the phase-modified correlation function~\eqref{eq:non-equilibrium_correlation_function}
%(\textcolor{blue}{17})
into the expression for the HOM excess noise~\eqref{eq:HOM_noise}
%(\textcolor{blue}{13})
, we obtain 
\begin{align}
\label{eq:hom_app}
    \Delta S_{\rm HOM} =& \frac{(q\nu|\Lambda|)^2}{(2\pi\alpha)^2}\int_{-\infty}^{\infty}dt\int_{-\infty}^{\infty}dt'~\left\{\left[\frac{\pi k_BT\alpha/v_F}{i\sinh(\pi k_BT (t-t'-i\frac{\alpha}{v_F}))}\right]^{4\delta} + \left[\frac{\pi k_BT\alpha/v_F}{i\sinh(\pi k_BT (-t+t'-i\frac{\alpha}{v_F}))}\right]^{4\delta}\right\}\notag \\
    &\times\left[e^{i\vartheta\Phi(t,t')} + e^{-i\vartheta\Phi(t,t')}-2\right].
\end{align}
We can now safely set $\nu=1$, $\delta=1/2$, $\vartheta=\pi$. By next assuming the hierarchy of scales $\alpha \ll w\ll \beta=(k_{\rm B}T)^{-1}$, i.e., at very small temperatures, we find that Eq.~\eqref{eq:hom_app} simplifies to
\begin{align}
\label{eq:delta_SHOM_electrons}
    \Delta S_{\rm HOM} = -\frac{(q|\Lambda|)^2}{(2\pi v_F)^2}\int_{-\infty}^{\infty}dt\int_{-\infty}^{\infty}dt'~\left[\frac{1}{(t-t'-i\alpha/v_F)^2} + \frac{1}{(-t+t'-i\alpha/v_F)^2}\right]\left[e^{i\pi\Phi(t,t')} + e^{-i\pi\Phi(t,t')}-2\right].
\end{align}
The factors $\exp[\pm i\pi \Phi]$ are handled with the identities
\begin{equation}
    \exp\left[\pm2i\arctan\left(\frac{t}{w}\right)\right]=-\frac{t\mp iw}{t\pm iw},
    \label{eq:arctan_identity}
\end{equation}
and by shifting $t,t'\to t,t'+ t_{\rm u}$, setting $\tau_d \equiv t_{\rm l} - t_{\rm u}$, we evaluate the integrals in Eq.~\eqref{eq:delta_SHOM_electrons} with the residue theorem. We then find
\begin{align}
    \Delta S_{\rm HOM} &= \frac{4(q|\Lambda|)^2}{v_F^2}\frac{\tau^2_d}{\tau^2_d+4w^2}.
\end{align}
In the same way, we obtain the HBT noise for single-electron injection as $\Delta S_{\text{HBT},j} = 2(q|\Lambda|/v_F)^2$. Thus, the standard HOM ratio \eqref{eq:HOM_Ratio}
%(\textcolor{blue}{14})
for electrons evaluates at zero temperatures to
\begin{align}
\label{eq:R_electron}
    \mathcal{R}_{\rm el}(\tau_d) = \frac{\tau^2_d}{\tau^2_d+4w^2}.
\end{align}
As the delay $\tau_d \to 0$, $\mathcal{R}_{\rm el} \to 0$, which can be interpreted as manifesting the Pauli principle at zero temperature (see, e.g, Refs.~\cite{Jonckheere2012Sep,Dubois2013Aug,Ferraro2013Nov}). Furthermore, as the electron state width $w\to 0$, $\mathcal{R}_{\rm el}(\tau_d)\to 1$, which can be interpreted as the overlap between the two injected electronic quantum states becoming vanishingly small for all $\tau_d$. 

At finite temperature, the calculation is more involved. To proceed in this case, we first note that it is always possible to re-arrange the HOM ratio as
\begin{subequations}
\begin{align}
    &\mathcal{R}=1-\frac{\mathcal{N}}{\mathcal{D}},\\
    &\mathcal{N}=\Delta S_{\mathrm{HOM}}-\sum_{j=\text{u,l}}\Delta S_{\mathrm{HBT},j},\\
    &\mathcal{D}=\sum_{j=\text{u,l}}\Delta S_{\mathrm{HBT},j}.
\end{align}
\end{subequations}
 By setting $\vartheta=\pi$ and $\delta=1/2$ in~\eqref{eq:hom_app} and using the identity~\eqref{eq:arctan_identity}, we find that $\mathcal{N}$ and $\mathcal{D}$ for electrons can be expressed as
\begin{subequations}
\begin{align}
    \mathcal{N}_{\rm el}&=\frac{(q|\Lambda|)^2}{(2\pi\alpha)^2}\int_{-\infty}^{+\infty}d\bar{t}\int_{-\infty}^{+\infty}d\tau\left[\chi(\bar{t})\chi^*(\bar{t}+\tau)\chi^*(\bar{t}+\tau_d)\chi(\bar{t}+\tau_d+\tau)+\mathrm{H.c.}\right]\pi^2\tau^2 G_{\mathrm{eq}}(\tau),\\
    \mathcal{D}_{\rm el}&=\frac{(q|\Lambda|)^2}{(2\pi\alpha)^2}\int_{-\infty}^{+\infty}d\bar{t}\int_{-\infty}^{+\infty}d\tau[i\chi(\bar{t})\chi^*(\bar{t}+\tau)-i\chi^*(\bar{t})\chi(\bar{t}+\tau)]\pi\tau G_{\mathrm{eq}}(\tau),
\end{align}
\label{eq:nd}
\end{subequations}
where the function
\begin{align}
 \chi(t)\equiv\sqrt{\frac{w}{\pi}}\,\frac{1}{t+iw},
\end{align}
and
\begin{equation}
    G_{\mathrm{eq}}(\tau)=\left[\frac{\pi k_BT\alpha/v_F}{i\sinh(\pi k_BT (\tau-i\frac{\alpha}{v_F}))}\right]^{2}
\end{equation}
is the equilibrium electronic (i.e., for $\delta=1/2$) correlation function. To arrive at Eq.~\eqref{eq:nd} we have adapted the derivation in Refs.~\cite{Ronetti2018Aug,Ferraro2018Dec} to the case of the non-periodic injection of a Leviton state with unit charge. Next, in Eq.~\eqref{eq:nd}, the integrals over $\bar{t}$ can be evaluated with the residue theorem, yielding
\begin{subequations}
\begin{align}
    \mathcal{N}_{\rm el}&=\frac{4w^2}{\left(\tau_d^2+4 w^2\right)}\times\frac{(q|\Lambda|)^2}{(2\pi\alpha)^2}\int_{-\infty}^{+\infty}d\tau \frac{4\pi\tau^2 w}{ \tau ^2+4 w^2}G_{\mathrm{eq}}(\tau),\\
    \mathcal{D}_{\rm el}&=\frac{(q|\Lambda|)^2}{(2\pi\alpha)^2}\int_{-\infty}^{+\infty}d\tau \frac{4 \pi \tau^2  w}{\tau ^2+4 w^2} G_{\mathrm{eq}}(\tau)\,.
\end{align}
\label{eq:nd2}
\end{subequations}
We now see that all temperature dependence in these expressions is encoded in $G_{\rm rq}(\tau)$ inside the integrals over $\tau$. However, these integrals cancel in the ratio $\mathcal{N}_{\rm el}/\mathcal{D}_{\rm el}$, leading to the final result
\begin{equation}
    \mathcal{R}_{\rm el}(\tau_d)=1-\frac{\mathcal{N}_{\rm el}}{\mathcal{D}_{\rm el}}=1-\frac{4w^2}{\tau_d^2+4w^2}=\frac{\tau_d^2}{\tau_d^2+4w^2},
\end{equation}
identical to the zero temperature result~\eqref{eq:R_electron}. We thus confirm that the electronic HOM ratio is fully independent of temperature~\cite{Dubois2013Aug}, which stands in stark contrast to the anyonic case~\eqref{eq:HOM_ratio_point}
%(\textcolor{blue}{20})
in the main text.
\end{onecolumngrid}


\begin{thebibliography}{83}
\bibitem [{nor()}]{normalization}%
  \BibitemOpen
  \href@noop {} {\bibinfo  {journal} {For correct units of noise, one should in principle replace the integration over $t$ in Eq.~\eqref{eq:HOM_noise} by $\int_{-\mathcal{T}_0/2}^{\mathcal{T}_0/2}dt/\mathcal{T}_0$, with $\mathcal{T}_0$ a very long averaging time. This replacement, however, has no impact on the dimensionless HOM ratio~\eqref{eq:HOM_Ratio}}\ }\BibitemShut {NoStop}%
\end{thebibliography}
\end{document}